\documentclass{kluwer}
\begin{document}
\begin{article}
\begin{opening}
\title{A multiwavelength investigation of unidentified EGRET sources }            

\author{P. \surname{Wallace}\email{pwallace@berry.edu}}
\institute{Berry College}
\author{S. \surname{Bloom}\email{sbloom@hsc.edu}}
\institute{Hampden-Sydney College}
\author{M. \surname{Lewis}\email{matthew.s.lewis@dartmouth.edu}}
\institute{Berry College; current address: Dartmouth College}   




\runningtitle{Unidentified EGRET sources}
\runningauthor{Wallace, Bloom, \& Lewis}



\begin{abstract}
 
Statistical studies indicate that the 271 point sources of high-energy gamma rays belong to two groups: a Galactic population and an isotropic extragalactic population. Many unidentified extragalactic sources are certainly blazars, and it is the intention of this work to uncover gamma-ray blazars missed by previous attempts. Until recently, searches for blazar counterparts to unidentified EGRET sources have focused on finding AGN that have 5-GHz radio flux densities $S_5$ near or above 1 Jy. However, the recent blazar identification of 3EG J2006-2321 ($S_5 = 260$ mJy) and other work suggest that careful studies of weaker flat-spectrum sources may be fruitful. In this spirit, error circles of 4 high-latitude unidentified EGRET sources have been searched for 5-GHz sources. The gamma-ray sources are 3EG J1133+0033, 3EG J1212+2304, 3EG J1222+2315, and 3EG J1227+4302. Within the error contours of each of the four sources are found 6 radio candidates; by observing the positions of the radio sources with the 0.81-m Tenagra II telescope it is determined that 14 of these 24 radio sources have optical counterparts with $R < 22$. Eight of these from two different EGRET sources have been observed in the $B$, $V$, and $R$ bands in more than one epoch and the analysis of these data is ongoing. Any sources that are found to be variable will be the objects of multi-epoch polarimetry studies. \end{abstract}




\end{opening}

\section{Introduction}

With the exception of the LMC, blazars constitute all identified high-energy gamma-ray sources with Galactic latitude $|b|>30^{\circ}$; in this region there are 30 unidentified sources (Hartman et al. 1999). Galaxy clusters may be responsible for a very small number of unidentified EGRET sources at high latitudes (Colafrancesco et al. 2001) but any high-energy gamma-ray emission from such clusters probably falls below EGRET's sensitivity (Reimer et al. 2003). It is also possible that the Galactic halo houses some of the persistent EGRET sources (Grenier et al. 2001), but to date this is only a conjecture. It remains that the only identified gamma-ray sources above 30 degrees Galactic latitude are blazars with flat-spectrum radio counterparts, and we suggest that there are others among the 30 unidentified sources with $|b|>30^{\circ}$.

The present article describes an effort to single out those high-latitude gamma-ray sources that are most likely to be blazars, and to suggest plausible optical and radio counterparts for them. Using NED, we have searched within the 95\% confidence contours of 4 EGRET  sources for 5-GHz radio sources; 24 such radio sources were found, and fields centered on these radio sources were subsequently observed with the 0.81-m Tenagra II telescope at Tenagra Observatory, with the goal of determining plausible optical counterparts to the radio sources, and thus to the gamma-ray sources. The $R$-magnitudes of these sources are determined using IRAF, and follow-up photometry has been performed on those for which optical counterparts were found.  

Our working assumption is that these 4 gamma-ray sources are blazars, and it is our goal to determine their most probable radio and optical counterparts. In cases where no plausible counterparts are found, it may be that the source is not a blazar, but a new class of high-energy gamma-ray source. This work complements that of Sowards-Emmerd, Romani, \& Michaelson (2003; hereafter SRM 2003), in which detailed spectroscopic information was compiled for a large number of northern gamma-ray counterpart sources and in which a number of plausible identifications were reported. Only one of the four present sources (J1133+0033) is found to have a compelling counterpart in SRM 2003. In addition, two of our sources (J1212+2304 and J1222+2315) are listed in SRM 2003 as having no plausible blazar counterpart at all. This work also complements that of Bloom et al. (2004), who has performed similar work on a different set of high-latitude unidentified sources.   

\section{Selection of gamma-ray and radio sources}

It is impractical to search the error boxes of all the unidentified sources for variable optical counterparts to 5-GHz radio sources. We therefore restrict our source sample in several ways. First, we look only at high Galactic latitudes, as sources with $|b|>30^{\circ}$ have low sky densities and are least susceptible to source confusion. Also, high-latitude EGRET sources are least likely to be of Galactic origin. Further, we look to those gamma-ray sources that display some evidence of being variable. We use the variability statistics $\tau$ and $I$, described in Torres et al. (2001). The $\tau$ statistic takes into account 145 marginal sources not listed in Hartman et al. (1999); the maximum likelihood fluxes of all 416 sources are then re-computed. From these data, each source was assigned a value of $\tau$, which is defined as $\tau = \sigma / \mu$, where $\sigma$ is the standard deviation of the individual viewing period fluxes and $\mu$ is their mean value. In this scheme one must pay attention not only to $\tau$ itself but to its error bars; we consider a source to be variable if its 68\% lower limit is equal or greater than unity. An assumption of the $I$-statistic is that on the time scales under consideration (weeks to years) pulsars are not variable. $I$ is then defined as $\mu_{\mathrm{source}}/\mu_{\mathrm{pulsars}}$, where $\mu = 100 \times \sigma \times F -1$; $F$ is a weighted mean of the individual EGRET fluxes for $E > 100$ MeV and $\sigma$ is their standard deviation. So for pulsars, $I = 1$; we consider a source to be variable if $I > 4$, corresponding to $> 6\sigma$ above pulsar ``variability''.

For an EGRET source to be included in this study, we require that at least $\tau$ or $I$ indicate that the source is variable. This restriction is meant to act as blunt instrument and is intended only to help us avoid those sources that show no evidence whatsoever of variability. (Also, it is important to realize that for these dim sources all means of determining gamma-ray variability are subject to significant uncertainties.) Cutting out sources with far southerly declinations leaves us with 12 candidates. Four of these have been searched for radio sources and optical counterparts have been sought.

Each of the four 95\% error boxes contain six 5-GHz radio sources that satisfy the flowing criteria (1) the flux density at this frequency is roughly 50 mJy or greater (this is near the limit of the Parkes and Green Bank 5-GHz surveys); (2) the spectral index $\alpha$, where $F_{\nu}\propto \nu^{\alpha}$, between 1.4 GHz and 5 GHz is greater than about -1.0; (3) the position is known to an uncertainty of about $5^{\prime\prime}$; (4) the radio source is well-separated from nearby sources (within astrometric uncertainty), and (5) the source's optical counterpart (if any) is not extended. In this way we have restricted our candidates to those with radio features like those of the population of EGRET blazars. In some cases one or two of these requirements was relaxed if the radio source fell particularly close to the gamma-ray centroid, or if other indicators were particularly favorable for a blazar identification. These radio sources, with their NED-resolved names, 5-GHz flux densities $S_5$, radio spectral index $\alpha$, and angular separation from the EGRET centroid are listed in Table 1. The listing is by increasing RA of the EGRET sources and then by increasing angular separation. The spectral indexes are determined by finding $S$ at frequencies of 1.4 and 5 GHz; in those cases where there is no known 1.4 GHz counterpart, no $\alpha$ is given.  

\section{Optical observations}

Optical observations of all 24 radio positions were made with the Tenagra 0.81-m telescope in Nogales, Arizona during 2003 April and May. Those with known $V$ magnitudes were observed for 300 or 600 s, depending on their brightness. Those with unknown magnitudes were observed for 1500 s each; the Tenagra 0.81-m reaches $R\sim21$ in this time under good sky conditions. In addition, Landolt standard stars (Landolt 1992) and faint calibration stars from the Hamburg Quasar Monitoring program (e. g., Borgeest \& Shramm 1994) were observed nightly to determine the photometric scale and atmospheric absorption. The data were reduced using the standard packages of the Image Reduction and Analysis Facility (IRAF). After the astrometric plate solutions were produced using Mira software and the USNO-A2.0 catalog, it was determined that 14 of the 24 had high-confidence optical counterparts (radio and optical sources within 2 seconds of arc). 

\section{Current results and future direction}

Results from initial $R$-band photometry of the optical candidates, performed from 8-22 May 2003 at Tenagra Observatory, are given in Table 2. ID's are from this work and from SRM 2003. Multi-epoch photometry in the $B$, $V$, and $R$ bands for J1133+0033 and J1212+2304 was performed at Tenagra from January-March 2004 and is currently being reduced and analyzed. In northern spring 2005 the remaining optical sources will be monitored for color and variability; those sources that show evidence of variability will be considered for optical polarimetry; any sources which show variability in flux and polarization will be considered the best candidates for association with gamma-ray blazars.

\section{References}

\noindent Bloom, S. D. et al., Proc Gam 2001, AIP 587, 648 (2001)

\noindent Bloom, S. D. et al., AJ, 128, 56 (2004)

\noindent Borgeest, U. \& Shramm, K. A\&A, 284, 764B (1994)

\noindent Hartman, R. C. et al., ApJS, 123, 79 (1999)

\noindent Landolt, A. U., AJ, 104, 340 (1992)

\noindent Reimer, O. et al., ApJ588, 155 (2003)  

\noindent Sowards-Emmerd, D. et al., ApJ, 590, 109 (2003)

\noindent Torres, D. et al., AN 322, 223 (2001)

\noindent Wallace, P. M. et al., ApJ 569, 36 (2002)

\section{Acknowledgements}

PMW thanks J. P. Halpern for spectroscopy on PKS 1130+008 \& PKS B1130+009. Also, thanks to Michael Schwartz and Paulo Holvorcem at Tenagra Observatories. This research has made use of the NASA/IPAC Extragalactic Database (NED) which is operated by the Jet Propulsion Laboratory, California Institute of Technology, under contract with the National Aeronautics and Space Administration.

\begin{center}

\vspace{0.1in}

\begin{tabular}{cclccc}
\hline
$\gamma$-ray &$\gamma$-ray &NED-Resolved &$S_5$ &$\alpha$ & Ang Sep\\
Src \#       &Src Name     &      Name   &(mJy) &           & (arcmin)\\
\hline
1. &3EG J1133+0033 &PKS B1130+008        &352 & -0.2 & 4.4  \\
   &               &PMN J1132+0033       &252 & -0.2 & 6.5  \\ 
   &               &LBQS 1130+003        &194 & ---  & 21.2 \\
   &               &PMN J1133+0059       &122 & -0.3 & 26.1 \\
   &               &4C -00.45            &451 & -0.8 & 67.1 \\
   &               &87GB 112551.8+005033 &74  & -0.5 & 68.7 \\
2. &3EG J1212+2304 &4C +23.29            &210 & -0.9 & 48.7 \\
   &               &87GB 120653.1+225922 &39  & ---  & 50.7 \\ 
   &               &4C +24.25            &67  & -0.9 & 53.5 \\
   &               &MG2 J121626+2335     &73  & -0.8 & 59.6 \\  
   &               &87GB 121037.6+221200 &42  & ---  & 69.4 \\
   &               &MG2 J120941+2220     &112 & -0.8 & 80.2?\\   
3. &3EG J1222+2315 &MG2 J122209+2311     &109 & -0.9 & 6.0  \\
   &               &MG2 J122402+2238     &99  & -0.1 & 42.0 \\
   &               &4C +22.35            &179 & -1.1 & 47.6 \\
   &               &4C +23.30            &123 & -1.1 & 49.4 \\
   &               &87GB 122307.4+240021 &44  & -0.7 & 52.1 \\
   &               &MG2 122443+2359      &98  & -1.0 & 52.9 \\   
4. &3EG J1227+4302 &B3 1224+428          &53  & -1.3 & 29.9 \\
   &               &B3 1224+439          &161 & -0.1 & 38.6 \\
   &               &B3 1222+438          &235 & -0.4 & 40.6 \\
   &               &B3 1225+442          &124 & -0.9 & 58.7 \\ 
   &               &B3 1219+438          &30  & -1.1 & 62.1 \\
   &               &3C 272               &360 & -1.2 & 62.6 \\ 
\hline
\end{tabular}
\vspace{0.1in}
{Table 1. Properties of radio candidates.}
\end{center}

\begin{center}

\begin{tabular}{clcccc}
\hline
$\gamma$-ray &NED-Resolved       &NED   & Exposure & Observed &$\sigma_R$ \\
Src \#       &Radio Name         &mag   & (s)      & $R$ mag  &           \\
\hline
1.        &PKS B1130+008         & 18.5 & 300  & 16.99 & 0.06 \\
          &PMN J1132+0033        &  --- & 1500 & ---   & ---  \\ 
          &LBQS 1130+003         & 18.7 & 300  & ---   & ---  \\
          &PMN J1133+0059        &  --- & 1500 & 19.87 & 0.30 \\         
          &4C -00.45             & 16.5 & 300  & 15.94 & 0.01 \\
          &87GB 112551.8+005033  &  --- & 1500 & 18.43 & 0.13 \\
 
2.        &4C +23.29             & 17.5 & 300  & 17.06 & 0.15 \\
          &87GB 120653.1+225922  &  --- & 1500 & 20.08 & 0.14 \\
          &4C +24.25             &  --- & 1500 & ---   & ---  \\
          &MG2 J121626+2335      &  --- & 1500 & 18.25 & 0.17 \\
          &87GB 121037.6+221200  &  --- & 1500 & 19.15 & 0.41 \\
          &MG2 J120941+1220      &  --- & 1500 & ---   & ---  \\

3.        &MG2 J122209+2311      &  --- & 1500 & 17.71 & 0.04 \\
          &MG2 122402+2238       &  --- & 1500 & ---   & ---  \\
          &4C +22.35             & 18.5 & 300  & 17.92 & 0.12 \\
          &4C +23.30             &  --- & 1500 & ---   & ---  \\
          &87GB 122307.4+240021  &  --- & 1500 & ---   & ---  \\
          &MG2 122443            &  --- & 1500 & ---   & ---  \\

4.        &B3 1224+428           &  --- & 1500 & ---   & ---  \\
          &B3 1224+439           &  --- & 1500 & 18.89 & 0.11 \\       
          &B3 1222+438           & 20.4 & 600  & 20.00 & 0.55 \\
          &B3 1225+442           & 18.2 & 300  & 18.64 & 0.25 \\          
          &B3 1219+438           &  --- & 1500 & ---   & ---  \\
          &3C 272                &  --- & 1500 & 20.08 & 0.50 \\
          
\hline
\end{tabular}
\vspace{0.1in}
{Table 2. Results of initial optical photometry.}
\end{center}

\end{article}
\end{document}